\documentclass[english,english]{article}
\usepackage[T1]{fontenc}
\usepackage[latin1]{inputenc}
\usepackage{babel}

\makeatletter

\providecommand{\LyX}{L\kern-.1667em\lower.25em\hbox{Y}\kern-.125emX\@}

\usepackage[T1]{fontenc}
\usepackage[latin1]{inputenc}
\usepackage{babel}

\makeatletter

\usepackage{amsfonts}
\usepackage{amscd}

\makeatother

\makeatother
\begin{document}

{\centering \textbf{\LARGE Realization of algebras with the help of
\( \star  \)-products}\LARGE \par}

\textbf{\bigskip}

\bigskip

{\centering Claudia Jambor%
\footnote{jambor@theorie.physik.uni-muenchen.de
}, Andreas Sykora%
\footnote{andreas@theorie.physik.uni-muenchen.de
}\par}

\bigskip

{\centering LMU München, Sektion Physik\par}

{\centering Theresienstr. 37, 80333 München\par}

{\centering \bigskip\par}

{\centering Max-Planck-Institut f\"ur Physik, \par}

{\centering F\"ohringer Ring 6, 80805 M\"unchen, Germany\par}

\begin{abstract}
We present a closed formula for a family of \( \star  \)-products
by replacing the partial derivatives in the Moyal-Weyl formula with
commuting vector fields. We show how to reproduce algebra relations
on commutative spaces with these \( \star  \)-products and give some
physically interesting examples of that procedure.
\end{abstract}
\newpage

\section{Introduction}

The general idea of a \( \star  \)-product is to introduce an invertible
map \( \Omega  \) from a function space to an algebra and then to
pull back the algebra product between the two operators \( \Omega (f),\, \Omega (g) \).
This results in the so called \( \star  \)-product between the two
functions on the function space

\begin{equation}
f\star _{M}g=\Omega ^{-1}(\Omega (f)\Omega (g))
\end{equation}
 which is both linear and associative.

The most commonly used \( \star  \)-product is the Moyal-Weyl \( \star  \)-product,
for which we take \( \mathbb {R}^{N} \) for the manifold \( M \)
and parametrise it by \( N \) coordinates \( x^{i} \). Further let
\( \theta ^{ij}=const.\, (i,j=1\cdots N) \) be a constant and antisymmetric
matrix, then \begin{equation}
\label{definiton_moyal_weyl_star_product}
f\star g=\sum ^{\infty }_{n=0}\frac{(ih)^{n}}{2^{n}n!}\theta ^{i_{1}j_{1}}\cdots \theta ^{i_{n}j_{n}}\partial _{i_{1}}\cdots \partial _{i_{n}}f\, \partial _{j_{1}}\cdots \partial _{j_{n}}g
\end{equation}
 is a \( \star  \)-product between the functions \( f \) and \( g \)
on \( \mathbb {R}^{N} \).

To define more general \( \star  \)-products we take \( M \) to
be an arbitrary (sufficiently smooth) finite dimensional manifold.
A \( \star  \)-product on \( M \) is an associative, \( \mathbb {C} \)-linear
product on the space of functions (with values in \( \mathbb {C} \))
on \( M \) given by \begin{equation}
\textstyle f\star g=fg+{\frac{h}{2}}B_{1}(f,g)+({\frac{h}{2}})^{2}B_{2}(f,g)+\cdots 
\end{equation}
 where \( f \) an \( g \) are two such functions and the \( B_{i} \)
are bi-differential operators on \( M \). The parameter \( h \)
is called deformation parameter. There is a natural gauge group acting
on \( \star  \)-products consisting of \( \mathbb {C} \)-linear
transformations on the space of functions\begin{equation}
f\rightarrow f+hD_{1}(f)+h^{2}D_{2}(f)+\cdots 
\end{equation}
 where the \( D_{i} \) are differential operators. They may be interpreted
as a generalizations of coordinate transformations. If \( D \) is
such a linear transformation it maps a \( \star  \)-product to a
new one by \begin{equation}
\label{linear_transformation_on_star_product}
f\star ^{\prime }g=D^{-1}(D(f)\star D(g)).
\end{equation}
 \( \star  \)-products related in such a way are called eqivalent.

If one expands equation (\ref{linear_transformation_on_star_product})
in \( h \) one gets that a linear transformation \( D \) acting
on \( \star  \) only affects the symmetric part of \( B_{1} \)\begin{equation}
B^{\prime }_{1}(f,g)=B_{1}(f,g)+D_{1}(fg)-fD_{1}(g)-D_{1}(f)g.
\end{equation}
 Since one can show that the symmetric part of \( B_{1} \) may be
cancelled by a linear transformation we may assume \( B_{1} \) to
be antisymmetric. Since \( \star  \) is associative, the commutator
\begin{equation}
\label{star-commutator}
[f\stackrel{\star }{,}g]=f\star g-g\star f=hB_{1}(f,g)+\cdots 
\end{equation}
 is a derivation with a Leibniz rule\begin{equation}
[f\star g\stackrel{\star }{,}h]=f\star [g\stackrel{\star }{,}h]+[f\stackrel{\star }{,}h]\star g.
\end{equation}
 Up to first order this means that the antisymmetric part of \( B_{1} \)
is a derivation with respect to both functions \( f \) and \( g \).
Additionally, since the \( \star  \)-product is associative, the
Jakobi-identity is fulfilled\begin{equation}
[f\stackrel{\star }{,}[g\stackrel{\star }{,}h]]+[h\stackrel{\star }{,}[f\stackrel{\star }{,}g]]+[g\stackrel{\star }{,}[h\stackrel{\star }{,}f]]=0,
\end{equation}
 which implies up to second order that \( B_{1}(f,g) \) is a Poisson
structure \( \{.\, ,\, .\} \) with \begin{equation}
\{f,\{g,h\}\}+\{h,\{f,g\}\}+\{g,\{h,f\}\}=0.
\end{equation}
 After a certain linear transformation we therefore can always write
for the \( \star  \)-product on a local patch of the manifold \begin{equation}
f\star g=fg+\frac{ih}{2}\Pi ^{ij}\partial _{i}f\, \partial _{j}g+\cdots 
\end{equation}
 with\begin{equation}
\label{Poisson_condition_local}
\Pi ^{il}\partial _{l}\Pi ^{jk}+\Pi ^{kl}\partial _{l}\Pi ^{ij}+\Pi ^{jl}\partial _{l}\Pi ^{ki}=0.
\end{equation}

Therefore we can classify \( \star  \)-products up to second order
by Poisson structures on the manifold. The other way around it is
even more interesting: if there is a manifold with a given Poisson
structure \( \{.,.\} \), it is possible to construct \( \star  \)-products
with\begin{equation}
f\star g=fg+\frac{ih}{2}\{f,g\}+\cdots .
\end{equation}
 This was first done for symplectic manifolds (manifolds with invertible
\( \Pi ^{ij} \)) in \cite{DeWilde:1983b,Fedosov:1994a}. In \cite{Kontsevich:1997vb}
a general construction for arbitrary Poisson manifolds is given (see
also \cite{Cattaneo:1999fm}). \\
 \bigskip

In the following we construct an associative (proof in section \ref{associativity})
\( \star  \)-product which provides a generalization of the Moyal-Weyl
product (\ref{definiton_moyal_weyl_star_product}) by replacing the
partial derivatives with commuting vector fields, since they have
similar properties. For this we give some guiding examples in two
dimensions to see that even in this simple cases interesting algebra
structures arise. Furthermore we propose a method to calculate \( \star  \)-products
that reproduce - even for more complicated deformed algebras - the
algebra relations in the according commutative spaces. With this method
we get closed explicit formulas for the \( \star  \)-product for
a generalization of \( M(so_{a}(n)) \), which might be useful for
theories like \cite{Dimitrijevic:2003wv,Dimitrijevic:2003pn} and
also for the physical relevant \( M(so_{q}(3)) \).

\section{Algebras and \protect\protect\( \star \protect \protect \)-products}

The space of functions on \( \mathbb {R}^{N} \) together with the
\( \star  \)-product (\ref{definiton_moyal_weyl_star_product}) forms
a representation of the algebra\begin{equation}
\mathcal{A}=\mathbb {C}\! <\! \hat{x}^{1},\cdots ,\hat{x}^{N}\! >/([\hat{x}^{i},\hat{x}^{j}]-ih\theta ^{ij}),
\end{equation}
 since \( [x^{i}\stackrel{\star }{,}x^{j}]=ih\theta ^{ij} \). It
is interesting to find \( \star  \)-products for other relation-defined
algebras like algebras with Lie algebra structures

\begin{equation}
[\hat{x}^{i},\hat{x}^{j}]=ihC^{ij}{}_{k}\hat{x}^{k},\hspace {1cm}h,C^{ij}{}_{k}\in \mathbb {C}
\end{equation}
 and quantum space structures \cite{Cerchiai:1998eg,Schmidke:1991mr,Wess:1999,Wess:1991vh}
with\begin{equation}
\hat{x}^{i}\hat{x}^{j}=qR^{ij}{}_{kl}\hat{x}^{k}\hat{x}^{l},\hspace {1cm}q=e^{h},R^{ij}{}_{kl}\in \mathbb {C}.
\end{equation}

Instead of considerering these special relations we discuss a more
general case. We assume that the algebra \( \mathcal{A} \) is generated
by \( N \) elements \( \hat{x}^{i} \) and relations

\begin{equation}
[\hat{x}^{i},\hat{x}^{j}]=\widetilde{\hat{c}}^{ij}(\hat{x})=ih\hat{c}^{ij}(\hat{x})
\end{equation}
 where we assume that the right hand side of this formula is containing
a parameter \( h \) and that it is getting small as this parameter
vanishes. The mathematically correct context would be a \( h \)-adic
expanded algebra \begin{equation}
\label{hadic_expanded_algabra}
\mathcal{A}=\mathbb {C}\! <\! \hat{x}^{1},\cdots ,\hat{x}^{N}\! >\! [[h]]\, /([\hat{x}^{i},\hat{x}^{j}]-ih\hat{c}^{ij}(\hat{x}))
\end{equation}
 where it is possible to work with formal power series in \( h \).
Note that this kind of algebras all fulfill the Poincare-Birkhoff-Witt
property since a re-ordering of two \( \hat{x}^{i} \) never affects
the polynomials of same order in \( h \). An algebra with Poicare-Birkhoff-Witt
property possesses also a basis of lexicographically ordered monomials.
For an algebra generated by two elements \( \hat{x} \) and \( \hat{y} \)
this means that the monomials \( \hat{x}^{n}\hat{y}^{m} \) constitute
a basis.

\section{\protect\protect\( \star \protect \protect \)-products with commuting
vector fields\label{associativity}}

Let \( X \) be a vector field. It is easy to show that\begin{equation}
X^{i}(x)\frac{\partial }{\partial x^{i}}\left( f(x)\, g(x)\right) =\left. (X^{i}(y)\frac{\partial }{\partial y^{i}}+X^{i}(z)\frac{\partial }{\partial z^{i}})\left( f(y)g(z)\right) \right| _{y\rightarrow x,z\rightarrow x}.
\end{equation}
 To write the rhs in a more compact way we introduce the following
notation\begin{equation}
X_{1}f_{1}g_{1}=\left. (X_{2}+X_{3})f_{2}g_{3}\right| _{2\rightarrow 1,3\rightarrow 1}.
\end{equation}
 With this we can derive a Leibniz rule\begin{eqnarray}
X^{l}_{1}f_{1}g_{1} & = & \left. (X_{2}+X_{3})^{l}f_{2}g_{3}\right| _{2\rightarrow 1,3\rightarrow 1}\\
P(X_{1})f_{1}g_{1} & = & \left. P(X_{2}+X_{3})f_{2}g_{3}\right| _{2\rightarrow 1,3\rightarrow 1}
\end{eqnarray}
 where \( P \) is a polynomial in the vector fields. The last equation
can also be written in the form \begin{equation}
\label{eq: vector}
P(X_{1})\left( \left. f_{2}g_{3}\right| _{2\rightarrow 1,3\rightarrow 1}\right) =\left. P(X_{2}+X_{3})f_{2}g_{3}\right| _{2\rightarrow 1,3\rightarrow 1}.
\end{equation}
 We now take \( n \) commuting vector fields \( X_{a}=X^{i}_{a}\partial _{i} \),
i.e.\( [X_{a},X_{b}]=0 \). In this case locally a coordinate system
\( y^{a}(x) \) with \( X_{a}=\partial _{y^{a}} \) always can be
found. Globally this does not have to be the case. Further let \( \sigma ^{ab} \)
be a constant matrix. Then we define a \( \star  \)-product via\begin{equation}
\label{def: vector_star_product}
\left. (f\star g)\right| _{1}:=\left. e^{\sigma ^{ab}X_{a2}X_{b3}}f_{2}g_{3}\right| _{2\rightarrow 1,3\rightarrow 1}.
\end{equation}
 To prove the associativity we calculate \begin{eqnarray}
\left. (f\star (g\star h))\right| _{1} & = & \left. e^{\sigma ^{ab}X_{a2}X_{b3}}f_{2}\left( \left. e^{\sigma ^{cd}X_{c4}X_{d5}}g_{4}h_{5}\right| _{4\rightarrow 3,5\rightarrow 3}\right) \right| _{2\rightarrow 1,3\rightarrow 1}\nonumber \\
 & = & \left. e^{\sigma ^{ab}X_{a2}(X_{b4}+X_{b5})}f_{2}e^{\sigma ^{cd}X_{c4}X_{d5}}g_{4}h_{5}\right| _{4\rightarrow 3,5\rightarrow 3,2\rightarrow 1,3\rightarrow 1}\\
 & = & \left. e^{\sigma ^{ab}X_{a1}X_{b2}+\sigma ^{ab}X_{a1}X_{b3}}e^{\sigma ^{cd}X_{c2}X_{d3}}f_{1}g_{2}h_{3}\right| _{2\rightarrow 1,3\rightarrow 1}\nonumber 
\end{eqnarray}
 and\begin{eqnarray}
\left. ((f\star g)\star h)\right| _{1} & = & \left. e^{\sigma ^{ab}X_{a1}X_{b2}}_{2}\left( \left. e^{\sigma ^{cd}X_{c3}X_{d4}}f_{3}g_{4}\right| _{3\rightarrow 1,4\rightarrow 1}\right) h_{2}\right| _{2\rightarrow 1}\nonumber \\
 & = & \left. e^{\sigma ^{ab}(X_{a3}+X_{a4})X_{b2}}e^{\sigma ^{cd}X_{c3}X_{d4}}f_{3}g_{4}h_{2}\right| _{3\rightarrow 1,4\rightarrow 1,2\rightarrow 1}\\
 & = & \left. e^{\sigma ^{ab}X_{a1}X_{b3}+\sigma ^{ab}X_{a2}X_{b3}}e^{\sigma ^{cd}X_{c1}X_{d2}}f_{1}g_{2}h_{3}\right| _{2\rightarrow 1,3\rightarrow 1},\nonumber 
\end{eqnarray}
 where in the second step we used the relation (\ref{eq: vector}).
The two expressions are equal since the vector fields commute according
to our assumption, so our new \( \star  \)-product is associative.

For an antisymmetric matrix \( \sigma  \) we get for the \( \star  \)-commutator
\begin{eqnarray}
[f\stackrel{\star }{,}g] & = & \left. \left( e^{\sigma ^{ab}X_{a1}X_{b2}}-e^{-\sigma ^{ab}X_{a1}X_{b2}}\right) f_{1}g_{2}\right| _{2\rightarrow 1}\nonumber \\
 & = & \left. 2\sinh (\sigma ^{ab}X_{a1}X_{b2})f_{1}g_{2}\right| _{2\rightarrow 1}.
\end{eqnarray}
 In the case of two vector fields, which we call \( X_{1}=X \) and
\( X_{2}=Y \), the explicit formula for \( \sigma ^{12}=h,\sigma ^{21}=0 \),
the asymmetric \( \star  \)-product, reads as \begin{equation}
\label{def: asym star}
f\star g=\sum ^{\infty }_{n=0}\frac{h^{n}}{n!}(X^{n}f)\, (Y^{n}g),
\end{equation}
 while for \( \sigma ^{12}=\frac{h}{2},\sigma ^{21}=-\frac{h}{2} \)
we get \begin{equation}
\label{def: antisym star}
f\star g=\sum ^{\infty }_{n=0}\frac{h^{n}}{2^{n}n!}\sum ^{n}_{i=0}(-1)^{i}{n\choose i}(X^{n-i}Y^{i}f)\, (X^{i}Y^{n-i}g),
\end{equation}
 which yields the antisymmetric \( \star  \)-product. Both \( \star  \)-products
have the same Poisson tensor \( \Pi ^{ij}=\sigma ^{ab}X^{i}_{a}\wedge Y^{j}_{b} \).

\section{Linear transformations}

If we have already a \( \star  \)-product, we have seen that we can
produce a new \( \star  \)-product just by a linear transformation
\( D \) on the space of functions (\ref{linear_transformation_on_star_product}).
Suppose that \( D \) is such an invertible operator and that its
expansion in derivatives starts with 1. We here assume, that \( D \)
is of the form\begin{eqnarray}
D=e^{\tau (X_{a})} & , & D^{-1}=e^{-\tau (X_{a})}
\end{eqnarray}
 where \( \tau  \) is a polynomial in the vector fields \( X_{a} \).
The \( X_{a} \) are still the commuting vector fields we used in
the last section. Then for the \( \star  \)-product (\ref{def: vector_star_product})
we see that \begin{eqnarray}
f\star 'g & = & D^{-1}(D(f)\star D(g))\nonumber \\
 & = & e^{-\tau (X_{a1})}\left( \left. e^{\sigma ^{ab}X_{a2}X_{b3}}e^{\tau (X_{a2})}f_{2}e^{\tau (X_{a3})}g_{3}\right| _{2\rightarrow 1,3\rightarrow 1}\right) \\
 &  & =\left. e^{-\tau (X_{a2}+X_{a3})+\sigma ^{ab}X_{a2}X_{b3}+\tau (X_{a2})+\tau (X_{a3})}f_{2}g_{3}\right| _{2\rightarrow 1,3\rightarrow 1}.\nonumber 
\end{eqnarray}
 For \( \tau  \) only quadratic in the \( X_{a} \) (note that \( \tau _{2}^{ab} \)
can be assumed to be symmetric, since according to the assumption
the vector fields commute) \begin{eqnarray}
\tau  & = & \tau _{1}^{a}X_{a}+\frac{1}{2}\tau _{2}^{ab}X_{a}X_{b}
\end{eqnarray}
 we have\begin{equation}
\tau (X_{a1})+\tau (X_{a2})-\tau (X_{a1}+X_{a2})=-\tau _{2}^{ab}X_{a1}X_{b2}
\end{equation}
 and the new \( \star  \)-product becomes\begin{equation}
f\star 'g=\left. e^{(\sigma ^{ab}-\tau ^{ab}_{2})X_{a1}X_{b2}}\textrm{ }f_{1}g_{2}\right| _{2\rightarrow 1}
\end{equation}
 We see that the antisymmetric \( \star  \)-product (\ref{def: antisym star})
and the asymmetric \( \star  \)-product (\ref{def: asym star}) are
related by a linear transformation in function space and therefore
are equivalent.

\section{Examples in two dimensions}

We choose two commuting vector fields and calculate the relations
of the \( \star  \)-product algebra with the asymmetric \( \star  \)-product
(\ref{def: asym star}).

\subsection*{\protect\protect\( X=x\partial _{x}\protect \protect \) , \protect\protect\( Y=\partial _{y}\protect \protect \)\label{example_so_a(n)}}

We take \( h=ia \) and get with that \begin{eqnarray}
x\star x=x^{2}, &  & y\star y=y^{2},\\
x\star y=xy+iax, &  & y\star x=xy,\nonumber 
\end{eqnarray}
 which yields \begin{equation}
[x\stackrel{\star }{,}y]=iax.
\end{equation}
 This is the algebra of two dimensional \( a \)-euclidean space \cite{Dimitrijevic:2003wv,Dimitrijevic:2003pn}.
This algebra is not isomorphic to the two dimensional Heisenberg algebra
\( [x,y]=ih \). \bigskip

\subsection*{\protect\protect\( X=(a+bx)\partial _{x}\protect \protect \) , \protect\protect\( Y=(c+dy)\partial _{y}\protect \protect \)}

For the general linear case we do analogous calculations and get \begin{equation}
[x\stackrel{\star }{,}y]=e^{bd}(y+\frac{c}{d})\star (x+\frac{a}{b}),
\end{equation}
 which is an interesting quantum space like structure. This example
includes the case of the Manin plane.

\bigskip

\subsection*{\protect\protect\( X=\frac{a}{\sqrt{x^{2}+y^{2}}}(x\partial _{x}+y\partial _{y})\protect \protect \)
, \protect\protect\( Y=x\partial _{y}-y\partial _{x}\protect \protect \)}

These are the derivatives \( \partial _{r} \) and \( \partial _{\theta } \)
of the coordinate transformation into spherical coordinates \( r \)
and \( \theta  \). We get \begin{equation}
[x\stackrel{\star }{,}y]=a\sqrt{x\star x+y\star y}.
\end{equation}

\bigskip

\subsection*{\protect\protect\( X=a(x\partial _{x}+y\partial _{y})\protect \protect \)
, \protect\protect\( Y=x\partial _{y}-y\partial _{x}\protect \protect \)}

This is a simplification of the previous case. We get \begin{equation}
[x\stackrel{\star }{,}y]=(\tan a)(x\star x+y\star y),
\end{equation}
 This algebra does not have the Poincare-Birkhoff-Witt property if
we would set \( \tan a=1 \). Therefore \( a \) has to be treated
as a formal parameter.

\section{Reconstruction of algebras\label{reconstruction}}

According to (\ref{star-commutator}) the \( \star  \)-commutator
of a \( \star  \)-product is a Poisson tensor up to first order \begin{equation}
[f\stackrel{\star }{,}g]=h\{f,g\}+\mathcal{O}(h^{2})=h\Pi (f,g)+\mathcal{O}(h^{2}),
\end{equation}
 where \( \Pi  \) is the Possion-bivector of the Poisson structure.
This we can get from the relation\begin{equation}
\label{Poisson_like_algebra_comm}
\{x^{i},x^{j}\}=\Pi ^{ij}=c^{ij}(x).
\end{equation}
 On the other hand, assume that this \( \star  \)-commutator reproduces
the algebra relations, then the rhs of the previous equation would
be a \( \star  \)-polynomial in the generators of the algebra, i.
e. \begin{equation}
[x^{i}\stackrel{\star }{,}x^{j}]=hc^{ij}_{\star }(x).
\end{equation}
 To calculate the leading order of \( c_{\star }(x) \) it is not
necessary to know the explicit form of the \( \star  \)-product,
since it always starts with the ordinary product of functions, so
we can substitute \( c_{\star }(x) \) by \( c(x) \), which is the
same as in (\ref{Poisson_like_algebra_comm}).

Now for the special case of the \( \star  \)-products (\ref{def: vector_star_product})
the Poisson structure is \begin{equation}
\Pi =\sigma ^{ab}X_{a}\wedge X_{b}.
\end{equation}
 If we are able to write a general the Poisson structure (\ref{Poisson_like_algebra_comm})
in this form, we can reconstruct the algebra relations with the help
of the \( \star  \)-product (\ref{def: vector_star_product}). We
now need a general method to find commutative vector fields with the
properties postulated in the previous sections. For this let \( f \)
be a function and \( X_{f}=\{f,\cdot \} \) the Hamiltonian vector
field associated to \( f \). Then the commutator of two such vector
fields is \begin{equation}
[X_{f},X_{g}]=X_{\{f,g\}},
\end{equation}
 due to the Jakobi identity of the Poisson bracket. If we can find
functions with \begin{eqnarray}
\{f_{i},g_{j}\}=\delta _{ij}, & \{f_{i},f_{j}\}=0, & \{g_{i},g_{j}\}=0,
\end{eqnarray}
 this implies that all commutators between the associated Hamiltonian
vector fields vanish. This functions need not to be unique. One can
deduce from the splitting theorem for Poisson manifolds \cite{Weinstein:1983}
that this is always possible in a neighborhood of a point if the rank
of the Poisson tensor is constant around this point. If we do not
want to find a \( \star  \)-product on \( \mathbb {R}^{N} \), but
a \( \star  \)-product with certain commutation relations, we can
reduce \( \mathbb {R}^{N} \) by the set of points where the rank
of the Poisson tensor jumps and we have a good chance to find functions
with the desired properties on the new manifold. In this case we can
write the Poisson tensor as \begin{equation}
\Pi =\sum _{i}X_{f_{i}}\wedge X_{g_{i}}.
\end{equation}
 In the following we present functions \( f_{i} \) and \( g_{i} \)
for Poisson tensors of several algebras and use the corresponding
Hamiltonian vector fields in the \( \star  \)-products (\ref{def: vector_star_product}).
We calculate the resulting algebra relations \( [.\stackrel{\star }{,}.] \)
and compare them to the original algebra relations.

\section{Examples for reconstruction of algebras}

The two examples we present here are just the most telling ones. The
other ones like for the \( q \)-deformed Heisenberg algebra\cite{Wess:1999}
, the Lie algebra \( so(3) \), the quantum spaces \( M(so_{q}(1,3)) \)\cite{Lorek:1997eh},
\( M(so_{q}(4)) \) \cite{Faddeev:1990ih,Ocampo:1996} and \( M(so_{q}(1,3)) \)
\cite{Kempf:1992pr,Kempf:1992re} can be found in \cite{Sykora_Diss,Jambor_Diss}.

\subsection{The quantum space \protect\protect\( M(so_{a}(n))\protect \protect \)}

Here we investigate the quantum space introduced in \cite{Majid:1994cy}.
Although this quantum space is covariant under the quantum group \( SO_{a}(n) \),
we never use this property. We took it because of its simple relations.
Further it has a nontrivial center. Since we are using the \( n \)-dimensional
generalisation introduced in \cite{Dimitrijevic:2003wv,Dimitrijevic:2003pn}
we will simply call it \( SO_{a}(n) \) covariant quantum space or
abreviated \( M(so_{a}(n)) \). The relations of this quantum space
are\begin{equation}
[\widehat{x}^{0},\widehat{x}^{i}]=ia\widehat{x}^{i}\; \; \; \mbox {for}\; \; \; i\neq 0,
\end{equation}
 with \( a \) a real number. The \( \hat{x}^{i} \) simply commute
with each other. In the following of the example Greek indices run
from \( 0 \) to \( n-1 \), whereas Latin indices run from \( 1 \)
to \( n-1 \). This algebra is a higher dimensional generalization
of the two dimensional algebra we calculated in section (\ref{example_so_a(n)}).
We take \( \mathbb {R}^{N} \) as the manifold and use coordinates
\( x^{0} \) and \( x^{i} \). With the two dimensional case in mind
we guess the following vector fields\begin{eqnarray}
X=iax^{i}\partial _{i}, &  & Y=\partial _{o}.
\end{eqnarray}
 For the asymmetric \( \star  \)-product (\ref{def: asym star})
we get with this vector fields\begin{equation}
[x^{i}\stackrel{\star }{,}x^{0}]=iax^{i}.
\end{equation}
 \bigskip

Even more useful is the generalization of the above algebra with the
new relations \begin{equation}
[\hat{x}^{\alpha },\hat{x}^{\beta }]=i(a^{\alpha }\hat{x}^{\beta }-a^{\beta }\hat{x}^{\alpha }),
\end{equation}
 where \( a^{\alpha } \) are now \( n \) different deformation parameters.
For this relations to be consistent the Jacobi identities have to
be fulfilled, which easily can be proven. Since the rhs of the relation
is linear and we are therefore dealing with a Lie algebra the Poisson
tensor associated with the algebra is just\begin{equation}
\{x^{\alpha },x^{\beta }\}=a^{\alpha }x^{\beta }-a^{\beta }x^{\alpha }.
\end{equation}
 If we want to find commuting vector fields that reproduce this Poisson
tensor we now follow the way outlined in section \ref{reconstruction}.
The rank of this matrix is \( 2 \). Therefore we have to find two
functions \( f,g \) fulfilling \( \{f,g\}=1 \). We make a guess
and define\begin{eqnarray}
f=a^{\alpha }x^{\alpha } &  & \tilde{x}^{\alpha }=x^{\alpha }-\frac{a^{\alpha }a^{\beta }}{a^{2}}x^{\beta }
\end{eqnarray}
 with \( a^{2}=a^{\alpha }a^{\alpha } \). These functions fulfill
commutation relations very similar to the special case of \( M(so_{a}(n)) \):\begin{eqnarray}
\{f,\tilde{x}^{\alpha }\}=a^{2}\tilde{x}^{\alpha } &  & \{\tilde{x}^{\alpha },\tilde{x}^{\beta }\}=0.
\end{eqnarray}
 If we define \( g=\frac{1}{a^{2}}ln\, \sqrt{\tilde{x}^{\alpha }\tilde{x}^{\alpha }} \)
we find\begin{equation}
\{f,g\}=1.
\end{equation}
 The commuting vector fields are now\begin{eqnarray}
X=\{f,\cdot \} & = & a^{2}x^{\beta }\partial _{\beta }-(a^{\alpha }x^{\alpha })a^{\beta }\partial _{\beta }\\
Y=\{\cdot ,g\} & = & -a^{-2}a^{\beta }\partial _{\beta }.
\end{eqnarray}
 In this case no singularities have shown up and the \( \star  \)-product
can be defined on whole \( \mathbb {R}^{n} \). Again we may use the
asymmetric \( \star  \)-product (\ref{def: asym star}) and see that
the algebra relations are reproduced.

\subsection{The quantum space \protect\protect\( M(so_{q}(3))\protect \protect \)}

Our second example is the \( \star  \)-product for the quantum space
\-\( M(so_{q}(3)) \), which is invariant under the quantum group
\( SO_{q}(3) \) \cite{Lorek:1997eh}. The algebra relations in the
basis adjusted to the quantum group terminology are \begin{eqnarray}
\hat{z}\hat{x}^{+}=q^{2}\hat{x}^{+}\hat{z}, & \hat{z}\hat{x}^{-}=q^{-2}\hat{x}^{-}\hat{z}, & [\hat{x}^{-},\hat{x}^{+}]=(q-q^{-1})\hat{z}^{2}.
\end{eqnarray}
 and therefore the Poisson brackets are \begin{eqnarray}
\{z,x^{+}\}=2zx^{+}, & \{z,x^{-}\}=-2zx^{-}, & \{x^{-},x^{+}\}=2z^{2}.
\end{eqnarray}
 For \( f=\frac{1}{2}\ln x^{-} \)and \( g=\ln z \) we get the desired
\( \{f,g\}=1 \), so the Hamiltonian vector fields become \begin{eqnarray}
X_{f}=z\partial _{z}+\frac{z^{2}}{x^{-}}\partial _{+}, &  & X_{g}=2(x^{+}\partial _{+}-x^{-}\partial _{-}).
\end{eqnarray}
 For the \( \star  \)-product we take \begin{eqnarray}
X=z\partial _{z}+\frac{\alpha z^{2}}{x^{-}}\partial _{+}, &  & Y=x^{+}\partial _{+}-x^{-}\partial _{-}.
\end{eqnarray}
 \bigskip

\noindent For the \textbf{asymmetric} \( \star  \)-product (\ref{def: asym star})
we get the following algebra relations \begin{eqnarray}
z\star x^{+} & = & e^{h}x^{+}\star z,\\
z\star x^{-} & = & e^{-h}x^{-}\star z,\\
\left[ x^{+}\stackrel{\star }{,}x^{-}\right]  & = & \frac{\alpha }{2}(e^{-2h}-1)z\star z.
\end{eqnarray}
 If we set \begin{eqnarray}
\textstyle q=e^{\frac{h}{2}}, &  & \alpha ={-\frac{2q^{2}}{q+q^{-1}}}
\end{eqnarray}
 this reproduces the original algebra relations.

\noindent \bigskip

\noindent For the \textbf{antisymmetric} \( \star  \)-product (\ref{def: antisym star})
the algebra relations become \begin{eqnarray}
z\star x^{+} & = & e^{h}x^{+}\star z,\\
z\star x^{-} & = & e^{-h}x^{-}\star z,\\
\left[ x^{+}\stackrel{\star }{,}x^{-}\right]  & = & -\frac{\alpha }{2}(e^{h}-e^{-h})z\star z.
\end{eqnarray}
 Again we get back the original algebra relations, if we set \begin{eqnarray}
\textstyle q=e^{\frac{h}{2}}, &  & \alpha ={-\frac{2}{q+q^{-1}}}.
\end{eqnarray}

\section*{Acknowledgements}

\noindent We want to thank Julius Wess for the possibility to work
on this topic and his steady support. Also we want to thank the MPI
and the LMU for their support.

\bibliographystyle{plain}
\bibliography{mainbib}

\end{document}